# Pseudo-Random Fluctuations, Stochastic Cooperativity and Burstiness in Dynamically Unstable High-Dimensional Biochemical Networks

**Simon Rosenfeld**

National Cancer Institute

EPN 3108, 6130 Executive Blvd, Bethesda, MD 20892

(301) 496-7748; sr212a@nih.gov

## Abstract

Two major approaches are known in the field of stochastic dynamics of intra-cellular biochemical networks. The first one places the focus of attention on the fact that many biochemical constituents vitally important for the network functionality may be present only in small quantities within the cell, and therefore the regulatory process is essentially discrete and prone to relatively big fluctuations. The second approach treats the regulatory process as essentially continuous. Stochasticity in such processes may occur due to bistability and oscillatory motion within the limit cycles.

The goal of this paper is to outline the third scenario of stochasticity in the regulatory process. This scenario is only conceivable in high-dimensional highly nonlinear systems, such as genetic regulatory networks (GRN). We focus attention on the fact that in the systems with size and link density of GRN (~25,000 and ~100, respectively) confluence of all the factors necessary for gene expression is a comparatively rare event, and only massive redundancy makes such events sufficiently frequent. An immediate consequence of this rareness is "burstiness" in mRNA and protein copy numbers, a well known experimentally observed effect. We introduce the concept of "stochastic cooperativity" and show that this phenomenon is a natural consequence of high dimensionality coupled with highly nonlinearity of a dynamical system. In mathematical terms, burstiness is associated with heavy-tailed probability distributions of stochastic processes describing the dynamics of the system. The sequence of stochastic cooperativity events allows for transition from continuous deterministic dynamics expressed in terms of ordinary differential equations (ODE) to discrete stochastic dynamics expressed in terms of Langevin and Fokker-Plank equations.

We demonstrate also that high-dimensional nonlinear systems, even in the absence of explicit mechanisms for suppressing inherent instability, may nevertheless reside in a state of stationary pseudo-random fluctuations which for all practical purposes may be regarded as stochastic process. This type of stochastic behavior is an inherent property of such systems and requires neither an external random force, nor highly specialized conditions of bistability.

## 1. Introduction

High-dimensional biochemical networks are the integral parts of intracellular organization. The most prominent roles in this organization belong to genetic regulatory networks [1] and protein-interaction networks [2]. Also, there are numerous other subsystems, such as metabolic [3] and glycomic networks [4], to name just a few. All these networks have several important features in common. First, they are highly diverse, i.e., contain numerous (up to tens of thousands) different *types* of molecules. Second, their dynamics is constrained by a highly structured, densely tangled intracellular environment. Third, their constituents are predominantly macro-molecules interacting in accordance with the laws of thermodynamics and chemical kinetics. Fourth, all these networks may be called “unsupervised” in the sense that they do not have an overlaying regulatory structure of a non-biochemical nature that would control the behavior of *each* individual molecule. Although the term “regulation” is frequently used in the description of cellular processes, its actual meaning is different from that in the systems control theory. In this theory, a controller is a device which is largely independent on the system under control. It has an ability to gather the information about the system in a non-invasive manner through sensors and signal exchange, but, importantly, this exchange has a different physical nature than the system-to-be-controlled. Notably, the regulatory signal produced by the controller and the way it directs the system are of a different physical nature than the functions of the system under control. Contrary to this picture, the intra- and inter-cellular regulations are of a biochemical nature themselves (e.g., protein signal transduction [5]); therefore, the subdivision of a system on the regulator and the subsystem-to-be-regulated is largely nominal. Logically, such a subdivision serves as a way of compartmentalizing a big biochemical system into relatively independent parts for the simplification of analysis. However, in biology this compartmentalization is rarely unambiguous, and it is never known for sure what regulates what. Another term frequently used in description of cellular dynamics is “machine” or “machinery.” Although these words are appropriate to express our fascination with the high level of organization of the cells and apparent purposefulness in their actions, yet the analogy of the machine fails to go far enough. The machines per se are being assembled in an environment different from that in which the machines are intended to work. There are always a design, a designer’s mind and a designer’s tools behind the creation of a machine; these tools are of a different physical nature than the operational modalities of the machine itself. On the contrary, the cellular machines are self-assembled; the resources of their creation are the same as the resources the machines are working with, that is the macromolecules and biochemical reactions. Similar reservations should be made regarding the concepts of “information”, “coding”, “signal” and other terms borrowed from the human technological and societal experiences. These analogies and metaphors are useful instruments assisting human imagination in dissecting cellular realities, but none of them should be taken too literally. The drawback of these instruments is that their indiscriminate usage obscures the fundamental fact that intracellular functionality is nothing else than a vast system of interconnecting biochemical reactions between billions of molecules belonging to tens of thousands of various molecular species. Therefore, studying general properties of such large biochemical systems is of primary importance for understanding functionality of the cell.

In this work, the focus of attention is placed on dynamical stability of the biochemical networks. First, we show that stringent requirements of dynamical stability have very little chance to be satisfied in the biochemical networks of sufficiently high order. Second, we show that a dynamically unstable system does not necessarily end up its existence through explosion or implosion, as prescribed by simple linear considerations. It is possible that such a system would reside in a dynamic state similar to a stationary or slowly evolving stochastic process. Third, we conjecture that the motion in a high-dimensional system of strongly interacting units inevitably includes a pattern of “bursting”, i.e., sporadic changes of the state variables in either positive or negative directions. In biology, burstiness is an experimentally observed phenomenon [6-9], and a number of theoretical works have been performed to understand its origins [10-13]. Two major facts are usually invoked to explain the burstiness. The first one is that some of the molecules critically important for the network functionality are present in the cell in small quantities, thus producing large random fluctuations downstream in the biochemical pathways. Another popular approach places the focus of attention on the oscillatory motion within limit cycles involving a number of simultaneous chemical reactions. The analysis presented in this paper points out to another possible scenario of bursting, in addition to the existing models. Unlike the two approaches just mentioned, the mechanism we consider does not require any special conditions for its realization. Rather, it is seen as a ubiquitous property of any high-dimensional highly nonlinear dynamical system, including biochemical networks. The mechanism of stochastic evolution proposed here allows for some experimentally verifiable predictions regarding global parameters characterizing the system.

## 2. Nonlinear Model and the State of Equilibrium

A natural basis for the description of chemical kinetics in a multidimensional network is the power-law formalism, also known under the name S-systems [14-17]. A useful property of S-systems is that S-functions are the “universal approximators,” i.e., have the capability of representing a wide range of nonlinear functions under mild restrictions on their regularity and differentiability. S-functions are found to be helpful in the analysis of genome-wide data, including those derived from microarray experiments [18]. However, the most important in the context of this work is the fact that in the vicinity of equilibrium any nonlinear dynamical system may be represented as an S-system [19]. Unlike mere linearization which replaces a nonlinear system by the topologically isomorphic linear one, the S-approximation still retains essential traits of nonlinearity but often is much easier to analyze.

Without loss of generality, the system of equations of chemical kinetics may be recast in the following form

$$\frac{dx_i}{dt} = F_i(x_1, ..., x_n) = \alpha_i \prod_{m=1}^{N} x_m^{p_{im}} - \beta_i \prod_{m=1}^{N} x_m^{q_{im}} \quad , \tag{1}$$

where $\alpha_i$ , $\beta_i$ are the rates of production and degradation, and $p_{im}$, $q_{im}$ are the stoichiometric coefficients in the direct and inverse reactions, respectively. Depending on the nature and complexity of the system under investigation, the quantities $\{x_i\}$, $i = 1, ..., N$ , may represent various biochemical constituents participating in the process, including individual biomolecules or their aggregates. In the context of GRN, these biochemical constituents may include proteins, mRNAs, DNA, and numerous transcription factors such as holoenzymes, promoters, repressors and others [20]. There is no unique way of representing the biochemical machinery in mathematical form: depending on the level of structural “granularity” and temporal resolution, the same process may be seen either as an individual chemical reaction or as a complex system of reactions. For example, the process that is commonly characterized as a singular event of “binding” of a protein to the regulatory site in reality is the sequence of events of enormous complexity involving a large number of rearrangements and supported by numerous transcriptional co-activators. Similarly, on a certain level of abstraction, the process of transcription may be seen as an individual biochemical reaction between RNA polymerase and DNA molecule, whereas a more detailed view reveals a complex “dance” involving hundreds of elemental steps, each representing a separate chemical reaction [21]. However, it is important to note that in principle, regardless the complexity of the system, its dynamics may be expressed in the form (1) with appropriate definitions of chemical constituents, $\{x_i\}$.

Simple algebra allows for transformation of equation (1) to a more universal and analytically tractable form

$$\frac{dz_i}{dt} = F_i(t) = \nu_i \left[ \mathrm{e}^{U_i(t)} - \mathrm{e}^{V_i(t)} \right] \tag{2}$$

where $U_i(t) = \sum_{m=1}^{N} P_{im} z_m$, $V_i(t) = \sum_{m=1}^{N} P_{im} z_m$ $P_{im} = p_{im} - \delta_{im}$, $Q_{im} = q_{im} - \delta_{im}$ (see Appendix A for technical details.) It is easy to see that the fixed point is located in the origin of coordinates and that the Jacobian matrix in its vicinity is simply

$$J_{im} = \nu_i (p_{im} - q_{im}) \tag{3}$$

No simplifications have been made for the derivation of (2-3). This means that these equations are quite general and may be always derived for any given sets of rates and stoichiometric coefficients.

## 3. Structure of the Solution in the Vicinity of Equilibrium

Equations (1,2) may be simultaneously viewed as equations of chemical kinetics derived from and governed by the laws of non-equilibrium thermodynamics, and also as the equations of a certain dynamical system, whether originating in chemistry or not. There is a fundamental difference between the dynamic equilibrium resulting from the conditions $dz_i/dt = 0, i = 1,...,N$, and the thermodynamic equilibrium expressed in the Acting Mass Law in chemical kinetics [22]. The latter assumes, in addition to the fact that the fixed point is the equilibrium point, existence of the detailed balance, i.e., full compensation of each chemical reaction by the inverse one. Since a system of biochemical reactions within the cell may be seen as a collection of spatially separated loops with mostly unidirectional flow of chemical constituents, we do not expect a *global* validity of the principle of detailed balance (PDB) in intracellular dynamics. A cascade of spatially separated processes supporting homeostatic equilibrium is a fundamental feature of living systems; it is very distinct from the state of “heat death” expressed in PDB. The principles governing the behavior of such systems (known as “dissipative structures”) have been formulated by Prigogine [23]. The most fundamental among these is the principle of minimal entropy production. It should be noted, however, that this principle is phenomenological by nature and cannot be directly derived from the equations of underlying dynamics [24]. Therefore, there are no first principles that would impose any limitations on the structure of the Jacobian matrix, $\mathbf{J}$, in the vicinity of the fixed point. This means in turn that $\mathbf{J}$ is just a matrix of general form, and there is no reason to expect that it has only the eigenvalues with negative real parts. Consequently, generally speaking, there are no reasons to assume that the macroscopic law of motion, i.e., $\mathbf{dx}/dt = \mathbf{F(x)}$, is stable. Although the assumption of stability is frequently introduced in the context of genetic regulation, in fact, it refers to a highly specific condition which is hardly possible in multidimensional systems with many thousands of independent governing parameters.

It is sometimes argued that this presumed instability contradicts the common perception of smooth behavior of living entities. It should be noted in this regard that there are several qualitatively similar concepts characterizing temporal evolution which are frequently used interchangeably in biology, but should be clearly distinguished in mathematical modeling. It is especially important to distinguish between stability and stationarity. In biology, it is often observed that certain states either do not change at all or evolve slowly. Such a behavior may be the result of a genuine dynamical stability of the system, but also it may be the result of averaging across the population of cells of stationary variations in an inherently unstable dynamical system. On a macro level, these two behaviors may look similar, and thus both deserve to be characterized as “stable.”

In this context, it is useful to recall some fundamental results pertaining to stability of nonlinear systems. According to the theorem by Lyapunov [25], the matrix $\mathbf{J}$ is stable if and only if the equation $\mathbf{J'V} + \mathbf{VJ} = -\mathbf{I}$ has a solution, $\mathbf{V}$, and this solution is a *positive definite* matrix [25]. If such a solution does exist then the criteria of stability are reduced to the sequence of inequalities

$$\left|V_{11}\right| > 0\,;\ \begin{vmatrix} V_{11} & V_{12} \\ V_{21} & V_{22} \end{vmatrix} > 0;...;\ \begin{vmatrix} V_{11} & V_{12} & ... & V_{1n} \\ V_{21} & V_{22} & ... & V_{2n} \\ ... & ... & ... & ... \\ V_{n1} & V_{n2} & ... & V_{nn} \end{vmatrix} > 0 \qquad (4)$$

Matrix $\mathbf{V}$ is a complicated function of all the stoichiometric coefficients and kinetic rates characterizing the network. Inequalities (4) would impose the set of very stringent constraints of high algebraic order on the structure of dynamically stable biochemical networks. Another classical approach to stability consists of the application of the Routh-Hurwitz criterion [25]. In this approach, one first calculates the characteristic polynomial of the Jacobian matrix; then one builds the sequence of the so-called Hurwitz determinants from its coefficients. The system is stable if and only if *all* the Hurwitz determinants are positive. Again, the Routh-Hurwitz criterion imposes the set of very complex constraints on global structure of a biochemical network. As argued above, apart from the PDB there are no other first principles and/or general laws governing stability of biochemical systems, and neither the Lyapunov nor Routh-Hurwitz criteria are the parts of PDB. As shown in a recent paper by the author [26], the Jacobian matrix of an arbitrary biochemical system may have comparable numbers of eigenvalues with negative and positive real parts. This property holds under widely varying assumptions regarding kinetic rates and stoichiometric coefficients. Therefore, generally, high-dimensional biochemical networks which are not purposefully designed to be stable (like in reactors for biochemical synthesis [27]) are reasonably presumed to be unstable. It is sometimes argued that the cell is not “a bag of molecules” but rather it is a complex “machine.” It is also hypothesized that in the process of evolution the “evolutionary pressure” moved the cell from one stable configuration to another. Not discarding such a possibility, one still has to recognize that the “evolutionary pressure” is not a law by itself but rather is a label for all the unknown laws which brought the living systems into existence. In this capacity, the “evolutionary pressure” is analogous to the “wisdom of nature” or even to the “intelligent design”; it is not an established fact, but rather an expression of hope that the mechanism providing stability will be discovered sooner or later. It is precisely the goal of this paper to show how an *apparent* stability may appear even in the dynamically unstable high-dimensional biochemical networks.

## 4. Stochastic Cooperativity

The term cooperativity is widely used in biology for describing multi-step joint actions of biomolecular constituents to produce a singular step in intracellular regulation [28, 29]. In intra-cellular regulatory dynamics, the term cooperativity reflects the fact that an individual act of gene expression is not possible until all the gene-specific co-activators are accumulated in the quantities sufficient for triggering the transcription machinery. In ODE terms, this means that $d\mathbf{z}/dt$ in (2) may noticeably deviate from zero only when the majority of arguments in $U_i$ and $V_i$ come to "cooperation" by simultaneously reaching vicinities of their respective maxima. The fact that such an event is not frequent in a multidimensional system is demonstrated by the following simple example. Let's assume that $x(t)$ and $y(t)$ are the Gaussian processes and consider the behavior of the process, $F(t) = \exp[\sigma x(t)] - \exp[\sigma y(t)]$. The pattern of this behavior is seen in Figure 1 whereby $F(t)$ fluctuates in the vicinity of zero most of the time, thus making no contribution to the variations of $z(t)$. However, sometimes $F(t)$ makes large excursions in either direction causing fast sporadic changes in $z(t)$. As shown in Figure 2, the distribution of $F(t)$ is approximately symmetric; this means that positive excursions are generally balanced by negative ones. This observation helps us to understand how it happens that an inherently unstable system nevertheless behaves decently and does not explode or implode as prescribed by its linear instability. In simplified terms, the reason is that sporadic deviations of concentrations in positive directions are followed, sooner or later, by the balancing responses in degradation, thus maintaining approximate equilibrium.

## 5. Probabilistic Structure of Burstiness

In order to envision a stochastic structure in the solution to equation (2) we make use of three fundamental results from the theory of stochastic processes, namely, (i) Central Limit Theorem (CLT) under the strong mixing conditions (SMC) [30] ; (ii) asymptotic distribution of level-crossings by stationary stochastic processes [31] and (iii) probabilistic structure of heavy-tailed (also known as bursting) processes [32]. We first notice that the arguments of $F_i(t)$ in (2) are combined into two linear forms,

$$U_i(t) = \sum_{m=1}^{N} P_{im} z_m \ , V_i(t) = \sum_{m=1}^{N} Q_{im} z_m \qquad (5)$$

in which only $n \ll N$ terms are non-zeros, where $n$ is the typical number of proteins facilitating gene expression; as mentioned above, this number may be of order from several dozens to hundreds. Generally, collections of transcription factors are gene-specific, and there is no explicit correlation between transcription rates and transcription stoichiometry. We therefore may view these collections as independently bootstrapped from the totality of the proteome. According to the CLT under the SMC, the sums of weakly dependent random variables are asymptotically normal. Note that according to the Lindeberg theorem [33], the terms $z_i(t)$ in $U_i(t)$ and $V_i(t)$ are not required to be identically distributed: boundedness of the second moments is sufficient for the validity of CLT. Validity of the SMC, as applied to $U_i(t)$ and $V_i(t)$ is easy to demonstrate by simulation. Importantly, the sums (5) are asymptotically normal even when the processes $z_i(t)$ are drastically non-Gaussian. Figures 3 and 4 provide an illustration of convergence to normality. In this example, individual time series $z_i(t)$ are selected drastically non-normal, namely lognormal, and average cross-correlation between $z_i(t)$ is selected on the level 0.15. Nevertheless, summation of only 80 series, $z_i(t)$, results in the stochastic processes, $U_i(t)$ and $V_i(t)$ which are fairly close to Gaussian. Thus, we conclude that $U_i(t)$ and $V_i(t)$ are approximately Gaussian. Their parameters are:

expectations

$$\mu_i(P) = \sum_{k}^{N} P_{im} E\left[z_m(t)\right]; \ \ \mu_i(Q) = \sum_{k}^{N} Q_{im} E\left[z_m(t)\right], \qquad (6)$$

variances

$$\begin{aligned} \theta_i^2(P) &= \mathrm{var}[U_i(t)] = \sum_{k,m}^{N,N} P_{im}\, P_{ik}\, \mathrm{cov}\left[z_m(t) z_k(t)\right] \\ \theta_i^2(Q) &= \mathrm{var}[V_i(t)] = \sum_{k,m}^{N,N} Q_{im}\, Q_{ik}\, \mathrm{cov}\left[z_m(t) z_k(t)\right], \end{aligned} \qquad (7)$$

and covariance

$$\Lambda_{ij}(P,Q) = cov\left[U_i(t),V_j(t)\right] = \sum_{k,m}^{N,N} P_{im}\,Q_{jk}\,\mathrm{cov}\left[z_m(t)z_k(t)\right] \,, \tag{8}$$

where $\mathrm{cov}\left(z_m z_k\right) = E\left(z_m z_k\right) - E\left(z_m\right)E\left(z_k\right)$

Since $U_i(t)$ and $V_i(t)$ are Gaussian, the processes $\exp[U_i(t)]$ and $\exp[V_i(t)]$ are lognormally distributed; their expectations and variances are, respectively,

$$\mathrm{M}_i = \exp\left[\mu_i(\cdot) + \frac{\theta_i^2(\cdot)}{2}\right];\ \Theta_i^2 = \exp\left[2\mu_i(\cdot) + \theta_i^2(\cdot)\right]\left\{\exp\left[\theta_i^2(\cdot)\right] - 1\right\} \tag{9}$$

where dot stands for $P$ or $Q$. The correlation coefficient between two exponentials is

$$\rho_{ij}(P,Q) = \left\{\exp\left[\Lambda_{ij}(P,Q)\right] - 1\right\}\left\{\left[\exp\left(\theta_i^2(P)\right) - 1\right]\left[\exp\left(\theta_j^2(Q)\right) - 1\right]\right\}^{-1/2} \tag{10}$$

The right-hand side in (2) is the difference of two lognormal random variables. Exact probabilistic distribution of this difference is unknown. We have found by simulation that these distributions may be reasonably well approximated by the Generalized Pareto Distribution (GPD)

$$G_{\xi,\beta}(x) = 1 - \left(1 + \xi x/\beta\right)^{-1/\xi},\ \xi \neq 0\,;\ G_{\xi,\beta}(x) = 1 - \exp\left(-x/\beta\right),\ \xi = 0 \tag{11}$$

More specifically, the *tail* distributions of

$$h_\sigma(x) = \left|\exp(\sigma x) - \exp(\sigma y)\right| \tag{12}$$

may be accurately represented through (11) with appropriately selected parameters $\xi = \xi(\sigma)$ and $\beta = \beta(\sigma)$. These dependencies are found in our work by simulation and are shown in Figure 5. Furthermore, very accurate analytical approximations are available for $\xi$ and $\beta$. It turns out that $\xi = \xi(\sigma)$ is nearly linear

$$\xi(\sigma) = u + v\sigma + w\sigma^2\,;\ u = \left(\pi/2 - 2\right)/\left(\pi - 2\right) = -0.376;\ v = 0.745;\ w = -0.088 \tag{13}$$

and $\beta = \beta(\sigma)$ is nearly exponential

$$\beta(\sigma) = \varphi/\left(p+q\right)\left[\exp(p\sigma) - \exp(-q\sigma)\right];\ p = 1.162;\ q = 2.753;\ \varphi = \sqrt{\pi}/\left(\pi - 2\right) = 1.553 \tag{14}$$

Although the primary goal for these approximations is to accurately capture only the tail distributions of $h_\sigma(x)$, within the interval $0.1 \le \sigma \le 2.75$, approximations (13-14) are quite satisfactory down to 0.1-quantile. Essentially, this means that GPD may serve as a very good representation for $h_\sigma(x)$ as a whole, not just for the tails. Figure 6 shows an example of fitting the GPD to $h_\sigma(x)$. The histogram in the right panel depicts empirical distribution of $h_\sigma(x)$ resulting from the Monte Carlo simulation; a solid envelope line belongs to the theoretical density of GPD with parameters $\xi(\sigma)$ and $\beta(\sigma)$ obtained from (13-14).

The fact that $h_\sigma(t)$ is representable through the "heavy-tailed" GPD is significant. As well known from the literature [32], stochastic processes with heavy-tailed distribution usually possess the property of "burstiness". This property means that a substantial amount of spectral energy of such processes is contained in the so-called "exceedances", i.e., in the short sporadic pulses beyond the certain predefined bounds. Figure 7 illustrates this concept. The top panel depicts the stochastic process

$$h_\sigma(t) = \exp[\sigma x(t)] - \exp[\sigma y(t)]; \tag{15}$$

where $x(t)$ and $y(t)$ are standardized independent Gaussian processes. The second panel shows the process of exceedances, $\tilde{h}_\sigma(t)$, defined as the part of $h_\sigma(t)$ jumping outside the interval $0.025 \le \text{Prob}(h_\sigma) \le 0.975$. Although $\tilde{h}_\sigma(t)$ spends only 5% of all the available time outside this interval, its variance is overwhelmingly greater than that of difference, $d_\sigma(t) = h_\sigma(t) - \tilde{h}_\sigma(t)$ (183 and 7698, respectively). On this basis, we may regard $d_\sigma(t)$ as a small background noise which only slightly distorts the strong signal provided by $\tilde{h}_\sigma(t)$. If we ignore this noise, then equation (15) acquires a familiar form of the Langevin equation

$$\frac{dz_i}{dt'} = F_i(t) = \nu_i \sum_{k=1}^{L_i} \mu_{ik} \delta(t - t_{ik}) \tag{16}$$

where $\mu_{ik}$ is the matrix of random Pareto-distributed amplitudes and $t_{ik}$ is the set of random point processes coinciding with the events of bursting.

Temporal locations of pulses, $t_{ik}$, are those corresponding to local maxima of $U_i(t)$ and $V_i(t)$. It is a well known result from the theory of level-crossing processes [31] that the sequence of such events in the interval $(0,t]$ asymptotically converges to a Poisson process with the parameter

$$\zeta = (1/2\pi)(1/\tau_0)\exp\left\{-a^2/\left[2\theta^2\right]\right\} \tag{17}$$

where $a \to \infty$ is the threshold of excursion; $\tau_0$ and $\theta^2$ are the correlation radius and variance of the generating Gaussian processes, respectively. On the basis of this asymptotic result, it may be reasonably assumed that for a finite, but sufficiently large $a$ the sequences, $t_{ik}$, may also form a set of Poisson processes with appropriately selected parameters. Figure 8 shows an example of simulation where the threshold is only slightly greater than the standard deviation, i.e., $a = 1.35\theta$. The QQ-plot and histogram of waiting times, $\Delta t_k = t_{k+1} - t_k$, clearly follow exponential distribution, which is an indication that the sequence $t_k$ forms a Poisson process. It is also worth mentioning that in this simulation the number of peaks in the interval $(0, T = 100000]$ predicted from the asymptotic theory, 703, is fairly close to the number of peaks actually found, 696. These two findings indicate that equation (17) is practically applicable under much milder conditions than $a \to \infty$.

## 6. Meaning of Stochastic Stability

Having the Langevin equation (16) in place, we may now derive the corresponding Fokker-Plank equation. For this purpose, we compute increments

$$z_i(T) - z_i(0) = \nu_i \int_0^T dt \left[ e^{U_i(t)} - e^{V_i(t)} \right] \tag{18}$$

over the period of time, $T$, encompassing many excursion events. Since $E\left[ z_i(T) - z_i(0) \right] = 0$, we have the following equation for the variances of increments.

$$\operatorname{var}\left[ z_i(T) - z_i(0) \right] = \nu_i^2 \int_0^T dt \int_0^T dt' E\left\{ \left[ e^{U_i(t)} - e^{V_i(t)} \right] \left[ e^{U_i(t')} - e^{V_i(t')} \right] \right\} \tag{19}$$

Denoting

$$R_i(|t - t'|) = E\left\{ \left[ e^{U_i(t)} - e^{V_i(t)} \right] \left[ e^{U_i(t')} - e^{V_i(t')} \right] \right\}, \tag{20}$$

and using the standard Dirichlet technique, we find

$$\operatorname{var}\left[ z_i(T) - z_i(0) \right] = 2\nu_i^2 \int_0^T R_i(\tau)(T - \tau)\, d\tau \tag{21}$$

By definition, the diffusion coefficient is

$$D_i = \frac{\partial \operatorname{var}\left[z_i(T) - z_i(0)\right]}{\partial T} = 2\nu_i^2 \int_0^T R_i(\tau) d\tau \tag{22}$$

Since the correlation radius is much smaller than the inter-event time, in the above integral $T$ may be extended to $\infty$. Therefore,

$$D_i = \nu_i^2 \int_0^\infty R_i(\tau) d\tau \tag{23}$$

Expression (20), after some inessential simplifications, may be reduced to

$$R_i(\tau) = \exp\left[2\lambda \sum_k E\left(z_k\right) + \lambda \sum_k \operatorname{var}\left(z_k\right)\right] \bullet \left\{\exp\left[\lambda \sum_k \operatorname{var}\left(z_k\right) r_k(\tau)\right] - 1\right\} \tag{24}$$

where $\lambda = n/N$ ; (see Appendix B for details.) In equation (24), $r_k(\tau)$ are the autocorrelation functions of individual series $z_k(t)$. Applying the saddle-point approximation to the integral (23) we come to the following expression for the diffusion coefficient (see Appendix C).

$$D_i = \frac{1}{2}\sqrt{\frac{\pi}{\lambda}} \nu_i^2 \exp\left(2\lambda z_G\right) \frac{T_G}{\Theta_G} \exp\left[2\lambda \Theta_G^2\right] \tag{25}$$

where $\Theta_G^2 = \sum_k \operatorname{var}\left(z_k\right)$ denotes the network-wide variance of fluctuations and

$T_G^2 = \Theta_G^2 \Big/ \left[\sum_k \operatorname{var}\left(z_k\right) / \tau_k^2\right]$ is the network-wide square of relaxation time. Equation (25) reveals important details of multidimensional diffusion in the biochemical networks. First, there is a collective random force created by the entire network characterized by the factor $\left(T_G/\Theta_G\right)\exp\left(2\lambda z_G + 2\lambda \Theta_G^2\right)$, the force acting uniformly upon all the individual constituents. But also there are individual motilities characterized by the factors $\nu_i^2$ that reflect the abilities of individual constituents to be excited by this collective random force. Equation (25) also means that all the rescaled constituent-specific concentrations, $Z_i(t) = z_i(t)\nu_i^{-1}$, have the same diffusion coefficient,

$$D_G = \frac{1}{2}\sqrt{\frac{\pi}{\lambda}} \frac{T_G}{\Theta_G} \exp\left[2\lambda\left(z_G + \Theta_G^2\right)\right]. \tag{26}$$

Equation (25) reflects general tendencies in the structure of fluctuations. First, the diffusion coefficient rapidly grows with the complexity of the network, i.e., with $\lambda$. Further, it is natural to assume that correlation times, $\tau_k$, are of the same order of magnitude as the corresponding times of chemical relaxation, $\nu_k^{-1}$, because both introduce characteristic time scales into the individual chemical reactions. Therefore, the entire system may be stratified in accordance with only one parameter, the kinetic rate $\nu_k$

Generally, the probabilistic state of a biochemical network may be characterized by joint distribution, $P(\mathbf{z},t)$ of all the chemical constituents which satisfies the multivariate Fokker-Plank equation (FPE) [34]. However, in light of the above simplifications, such a detailed description would be redundant. Instead, we introduce a collection of $N$ identical univariate probability distributions, $P(Z,t)$, where $Z$ is any of the $Z_i = z_i\nu_i^{-1}$, each satisfying *the same* FPE with the coefficient of diffusion (24). Implicitly, equations in this system are not independent because the individual diffusion coefficients depend on the state of the entire network; however this moment-level dependence evolves in a much slower time scale then the correlation radius of fluctuations.

There are two important consequences of the fact that all the $Z_i(t) = z_i(t)\nu_i^{-1}$ satisfy the same FPE. First, it means that variances of chemical constituents, $\mathrm{var}(z_i)$, are directly proportional to the squares of corresponding kinetic rates. Since $z_i = \ln(y_i)$, we conclude that $\mathrm{var}\left[\ln(y_i)\right] \sim \nu_i^2$. This is a testable property of all the large scale biochemical networks and may serve as a basis for experimental validation. It gives some meaning to the experimentally observable fact that slow processes have low variances, whereas rapid fluctuations have high variances thus overwhelmingly contributing to unpredictable “noise" [35]**.** Since $\nu_i$ is the only temporal scaling constituent-specific parameter in the network, it is natural to surmise that the times of correlation, $\tau_i$, are directly proportional to the corresponding times of chemical relaxation, $\nu_i^{-1}$. This is another macroscopically observable property suitable for experimental validation.

As is well known, there are two major types of problems associated with FPE; the stationary problem and the initial value problem [34]. None of these approaches seems to be quite adequate for intracellular networks. Stationary approach is only meaningful when there are permanent influx and outflow of chemical constituents to and from the system, and, in addition, the time frame available for observation is substantially greater than *all* the $\nu_i^{-1}$. An alternative approach, i.e., the initial value problem requires knowledge of the initial state, $P(\mathbf{z}, t_0)$. If such information is available, then the solution to FPE produces a non-stationary process known as multidimensional random walk. Between these two polar cases, there is a vast variety of possible intermediate conditions which can be roughly categorized by the characteristic times, $\nu_i^{-1}$. As known from the biochemical observations, these characteristic times may cover wide spectrum of fluctuations ranging from milliseconds to many days [36]. The FPE framework proposed in this paper has an advantage of flexibility allowing one to treat each process individually, from stationary approach for the rapidly evolving constituents to the initial value approach for slow processes. Although not being trivially simple, the FPE methodology is nevertheless much simpler than tackling the original ODE system (1). Importantly, the stochastic model introduced in this paper does not depend on arbitrary assumptions regarding extraneous random noise of unknown origin. It emerges solely as a consequence of the deterministic equations of chemical kinetics (1) thus better conforming to the Ockham's *lex parsimoniae* principle.

Due to random partitioning and stochasticity of transcription initiation, the initial condition for the system of equations (21) may be considered as random. Starting with these initial conditions, the system is predominantly driven by the sequence of sporadic events of stochastic cooperativity. Although each such event produces a noticeable momentary shift in the system's evolution, the multitude of such events makes its overall behavior quite smooth; this behavior is illustrated in Figure 7, bottom panel. In principle, solution of the FPE with initial conditions is a non-stationary process in which individual variances of constituent-specific concentrations grow linearly with time. However, this non-stationarity is only noticeable in large time scales and reflects slow evolution of the system as a whole. Linear growth of variances with time reflects a progressive desynchronization of the biochemical interactions. This process may be viewed as slow "aging" of the system, whereas the sequence of stochastic cooperativity events itself looks more like the system's current "business as usual."

## 7. Discussion

In this paper, the Pareto representation of exceedances has been derived from the fact that $U_i(t)$ and $V_i(t)$ are Gaussian processes, and, therefore, $\exp[U_i(t)]$ and $\exp[V_i(t)]$ are lognormally distributed. We have justified the normality of $U_i(t)$ and $V_i(t)$ by the CLT. This assumption, however, served the only goal to simplify the analysis and may be substantially relaxed at the expense of increased complexity of calculations. Conceptually, all the major ideas leading to the notion of stochastic cooperativity would stay in place even without transition to asymptotic normality. Let's assume again, as we did in the examples in Figures 3 - 4, that $\{U_i(t), V_i(t)\} = \sum_m \{P_{im}, Q_{im}\} z_m$, where $z_i(t)$ are lognormal processes. This time, however, we do not assume that the number of non-zero elements in these sums is sufficiently large to equate the distributions of sums to their asymptotic limits. This would reflect the situation when the number of transcription factors is comparatively small (quite a rare case, as far as gene expression is concerned!) Generally, exact analytical expressions for the distributions of sums of lognormals are unknown, but there is a consensus in the literature that they may be accurately modeled as lognormally-distributed themselves [37]. We have performed the simulation for studying probabilistic structure of the exceedances with lognormal $U_i(t)$ and $V_i(t)$. It is rather remarkable that the GPD turns out to be a good approximation in this case as well, with the only reservation that simple parameterizations (12-13) are no longer valid and should be replaced by more complex ones.

The property of stochastic cooperativity is not limited to a special form of dynamical systems introduced through S-functions. There is a much wider class of nonlinear systems where the above outlined approach may be applied as well. Let, for example, $\{F_i(x)\}$ be a vector of positive monotonic functions with all $F_i'(x) > 0$ in $-\infty < x < \infty$. Let us also assume that $\mathbf{P}$ and $\mathbf{Q}$ are two matrices of all positive elements independently drawn from the same distribution. If the parameter $\sigma$ is sufficiently large, then the dynamics of the system

$$\frac{d\mathbf{x}}{dt} = \mathbf{F}(\sigma\mathbf{P}\mathbf{x}) - \mathbf{F}(\sigma\mathbf{Q}\mathbf{x}) \qquad (27)$$

will possess the properties of stochastic cooperativity and burstiness. The simplest way to envision this fact is to recall that, in accordance to the above mentioned property of S-functions to be "universal approximators", any nonlinear system may be represented through S-functions in the vicinity of the fixed point. This and other generalizations are currently under development and the results are to be published elsewhere.

There is a vast literature devoted to modeling the intracellular dynamics as a sequence of discrete events, famously exemplified by the Boolean Networks [38], Cellular Automata [39] and Artificial Genome [40], to name just a few (see [41] for comprehensive review.) Generally, the discrete and continuous descriptions largely exist as two separate, methodologically alternative realms. In the discrete dynamics, it is difficult, if possible at all, to realistically incorporate biochemical processes; differential equations of chemical kinetics is the only way to explicitly utilize the laws of molecular interactions [42]. On the other hand, when delving into intricate details of the trajectories in the state space with dimension in thousands, it is easy to loose a big picture and overall logic of events. Only the abstract discrete models seem to be able to reveal this logic. The model proposed in this work may serve as a natural bridge between the two approaches. Although in the time scale of the lifecycle of the cell the chemical reactions are essentially continuous processes, the sequence of events of stochastic cooperativity is essentially discrete.

## 8. Summary

We have outlined the mechanism by which a multidimensional autonomous nonlinear system, despite being dynamically unstable, nevertheless may be stationary, that is, may reside in a state of stochastic fluctuations obeying the probabilistic laws of random walk. Importantly, in this mechanism, the transition from the deterministic to probabilistic laws of motion does not require any artificial assumptions regarding the presence of extraneous random noise of unknown origin; stochastic-like behavior is produced by the system itself. An important role in forming this type of fluctuative motion belongs to inherent burstiness of the system associated with the events of stochastic cooperativity. Unlike classical Langevin approach, macroscopic laws of motion of the system are not required to be dynamically stable. The overall properties of stochastic cooperativity and burstiness are largely independent of the structure of the Jacobian's eigenvalues in the vicinity of a fixed point.

In this work, we have selected the S-systems to be an example of a nonlinear system. At least three motivations justify this selection. First, the S-systems are structured after the equations of chemical kinetics, thus being a natural tool for description of high-dimensional biochemical networks. Second, many other nonlinear systems may be represented through the S-systems in the vicinity of a fixed point. Third, despite generality, the S-systems have an advantage of being analytically tractable. However, many results regarding stochastic cooperativity and burstiness may be readily extended to other multidimensional nonlinear systems. In such system, short pulses during the events of stochastic cooperativity may be described in terms of "shot" noise with subsequent derivation of the Fokker-Plank equation. As proposed in this paper, it is possible to indicate some general experimentally verifiable predictions regarding the behavior of this type of system, such as distribution of intensities of fluctuations and distribution of temporal autocorrelations among individual units of the system.

## Appendix A. Derivation of the Equations (2-3)

Following the standard procedures in nonlinear dynamics [43], we first search for the state of dynamical equilibrium, $\{x_m^0\}$, commonly referred to as a "fixed" point, i.e., the point where all the time derivatives turn to zero, and which therefore satisfy the equations

$$\alpha_i \prod_{m=1}^{N} \left[ x_m^0 \right]^{p_{im}} = \beta_i \prod_{m=1}^{N} \left[ x_m^0 \right]^{q_{im}} \tag{A1}$$

Taking logarithm of both sides and solving the linear equations, we obtain the vector of solutions

$$x_i^0 = \exp\left[ \sum_{m=1}^{N} (p_{im} - q_{im})^{-1} \ln\left( \frac{\beta_m}{\alpha_m} \right) \right] \tag{A2}$$

Note that stoichiometric coefficients, $p_{im}$ and $q_{im}$ cannot be identical *in all* the direct and inverse reactions simultaneously, therefore, the matrix $p_{im} - q_{im}$ is always invertible.

It is convenient to introduce relative quantities, $y_i = x_i / x_i^0$ and then, after denoting

$$\mathbf{U = (p - I)(p - q)^{-1}} \;;\; \mathbf{V = (q - I)(p - q)^{-1}} \,, \tag{A3}$$

to obtain the equations

$$\frac{dy_i}{dt} = A_i \prod_{m=1}^{N} y_m^{p_{im}} - B_i \prod_{m=1}^{N} y_m^{q_{im}} \quad , \tag{A4}$$

where

$$A_i = \alpha_i \exp\left\{ \sum_{m=1}^{N} U_{im} \ln\left( \frac{\beta_m}{\alpha_m} \right) \right\} \;;\; B_i = \beta_i \exp\left\{ \sum_{m=1}^{N} V_{im} \ln\left( \frac{\beta_m}{\alpha_m} \right) \right\} \tag{A5}$$

Since we are interested only in positive solutions, we replace $y_i = \exp(z_i)$ and obtain

$$\frac{dz_i}{dt} = A_i \exp(\sum_{m=1}^{N} P_{im} z_m) - B_i \exp(\sum_{m=1}^{N} Q_{im} z_m) \,, \tag{A6}$$

where $P_{im} = p_{im} - \delta_{im}$, and $Q_{im} = q_{im} - \delta_{im}$.

We note further that

$$\frac{B_i}{A_i} = \frac{\beta_i}{\alpha_i} \exp\left\{ \sum_{m=1}^{N} (V_{im} - U_{im}) \ln\left( \frac{\beta_m}{\alpha_m} \right) \right\} \qquad \text{(A7)}$$

and, because $\mathbf{V - U = (q - p)(p - q)^{-1} = -I}$, we find that $B_i = A_i$, therefore

$$\frac{dz_i}{dt} = A_i \left\{ \exp\left(\sum_{m=1}^{N} P_{im} z_m\right) - \exp\left(\sum_{m=1}^{N} Q_{im} z_m\right) \right\} \qquad \text{(A8)}$$

After introducing a more appropriate time scale $t\bar{A} = t'$, where $\bar{A} = N^{-1} \sum_{i=1}^{N} A_i$, we rewrite equation (A8) as

$$\frac{dz_i}{dt'} = F_i(z) = \nu_i \left\{ \exp\left(\sum_{m=1}^{N} P_{im} z_m\right) - \exp\left(\sum_{m=1}^{N} Q_{im} z_m\right) \right\}, \qquad \text{(A9)}$$

where $\nu_i = A_i / \bar{A}$ with an important property that $\bar{\nu}_i = 1$. It is easy to see that now the fixed point is located in the origin of coordinates and that the Jacobian matrix in the vicinity of this point is

$$J_{im} = \nu_i (p_{im} - q_{im}) \qquad \text{(A10)}$$

## Appendix B. Derivation of the Autocorrelation Function

We start with two inessential simplifications. First, if the number of components in the sums $U_i(t) = \sum_{m=1}^{N} P_{im} z_m$, $V_i(t) = \sum_{m=1}^{N} P_{im} z_m$ is sufficiently large then we may ignore the differences between $\mathbf{P}, \mathbf{Q}$ and $\mathbf{p}, \mathbf{q}$, respectively, in the expressions (7) for variances $\theta_i^2(P)$ and $\theta_i^2(Q)$. Another simplification is based on the computationally established fact that the coefficients of variation of $\theta_i^2(P)$ and $\theta_i^2(Q)$ across the genes are much smaller than that of $\mathrm{var}(z_i)$. Therefore, we can replace $\theta_i^2(P)$ and $\theta_i^2(Q)$ by their averages across the genes (see [26] for more detail). After these simplifications

$$R_i(\tau) = \exp\{2E(U_i) + \mathrm{var}[U]\}\left\{\exp\{\mathrm{cov}[U(0), U(\tau)]\} - \exp\{\mathrm{cov}[U(0), V(\tau)]\}\right\} \qquad \text{(B1)}$$

We have further,

$$\mathrm{var}(U \mid P) = \sum_{k,m} p_k p_m \,\mathrm{cov}\left(z_k, z_m\right) \qquad \text{(B2)}$$

and

$$\mathrm{var}(U) = \lambda \sum_k \mathrm{var}\left(z_k\right) + \lambda^2 \sum_{k,m} \mathrm{cov}\left(z_k, z_m\right) \tag{B3}$$

Similarly,

$$\mathrm{cov}\left\{\left[U(0), U(\tau)\right] \mid P\right\} = \sum_{k,m} p_k p_m \,\mathrm{cov}\left[z_k(0), z_m(\tau)\right] \tag{B4}$$

and

$$\mathrm{cov}\left[U(0), U(\tau)\right] = \lambda \sum_k \mathrm{cov}\left[z_k(0), z_k(\tau)\right] + \lambda^2 \sum_{k,m} \mathrm{cov}\left[z_k(0), z_m(\tau)\right] \tag{B5}$$

At last,

$$\mathrm{cov}\left\{\left[U(0), V(\tau)\right] \mid P, Q\right\} = \sum_{k,m} p_k q_m \,\mathrm{cov}\left[z_k(0), z_m(\tau)\right] \tag{B6}$$

and

$$\mathrm{cov}\left[U(0), V(\tau)\right] = \lambda^2 \sum_{k,m} \mathrm{cov}\left[z_k(0), z_m(\tau)\right] \tag{B7}$$

Putting everything together

$$\begin{aligned} R_i(\tau) = \exp\left\{2E(U_i) + \left[\lambda \sum_k \mathrm{var}\left(z_k\right) + \lambda^2 \sum_{k,m} \mathrm{cov}\left(z_k, z_m\right) + \lambda^2 \sum_{k,m} \mathrm{cov}\left(z_k(0), z_m(\tau)\right)\right]\right\} \bullet \\ \bullet \left\{\exp\left\{\lambda \sum_k \mathrm{cov}\left[z_k(0), z_k(\tau)\right]\right\} - 1\right\} \end{aligned} \tag{B8}$$

The terms $\lambda^2 \sum_{k,m} \mathrm{cov}\left(z_k, z_m\right) + \lambda^2 \sum_{k,m} \mathrm{cov}\left(z_k(0), z_m(\tau)\right)$ are small compared to $\lambda \sum_k \mathrm{var}\left(z_k\right)$; first, because $\lambda \ll 1$, and second, because the double sums here are of the same order of magnitude as the sums of variances. The latter claim is supported by simulation. We therefore simplify (B8) to

$$R_i(\tau) = \exp\left[2E(U_i) + \lambda \sum_k \mathrm{var}\left(z_k\right)\right] \bullet \left\{\exp\left[\lambda \sum_k \mathrm{var}\left(z_k\right) r_k(\tau)\right] - 1\right\} \tag{B9}$$

## Appendix C. Derivation of the Diffusion Coefficient Using the Saddle-Point Approximation

Let $R(\tau)$ be a decreasing function of $\tau$ such that: $R(0) \gg 1; R(\infty) = 0;$ $R'(0) = 0$. Then

$$J = \int_0^\infty \left\{\exp\left[R(\tau)\right] - 1\right\} d\tau \approx \int_0^\infty \left\{\exp\left[R(0) + \tau R'(0) + \frac{\tau^2}{2} R''(0)\right] - 1\right\} d\tau \tag{C1}$$

Denoting $\sigma^2 = -1/R''(0)$, we find that $\exp\left[R(0) - \tau^2/\left(2\sigma^2\right)\right]$ is a good representation of the integrand in (C1) both in the vicinity of zero and at infinity. Therefore,

$$J = \exp R(0) \int_0^\infty \exp\left[-\frac{\tau^2}{2\sigma^2}\right] d\tau = \frac{\sqrt{\pi}}{2}\sigma e^{R(0)} = \frac{1}{2}\sqrt{\frac{\pi}{-R''(0)}} e^{R(0)} \tag{C2}$$

Introducing $\Lambda_i = \exp\left[2E(U_i) + \lambda\sum_k \operatorname{var}\left(z_k\right)\right]$, and $R(\tau) = \lambda\sum_k \operatorname{var}\left(z_k\right) r_k(\tau)$

we obtain

$$R_i(\tau) = \Lambda_i \bullet \exp\left[R(\tau) - 1\right] \tag{C3}$$

Denoting $\tau_k^{-2} = -\ddot{r}_k(0)$, we get $\sigma^{-2} = \lambda\sum_k \left[\operatorname{var}\left(z_k\right)/\tau_k^2\right]$

Therefore,

$$D_i = \nu_i^2 \exp\left[2E(U_i) + \lambda\sum_k \operatorname{var}\left(z_k\right)\right] \frac{\sqrt{\pi}}{2} \frac{\exp\left[\lambda\sum_k \operatorname{var}\left(z_k\right)\right]}{\sqrt{\lambda\sum_k \frac{\operatorname{var}\left(z_k\right)}{\tau_k^2}}} \tag{C4}$$

Introducing the parameters

$$\Theta_G^2 = \sum_k \operatorname{var}\left(z_k\right);\ T_G^2 = \sum_k \operatorname{var}\left(z_k\right) \Big/ \sum_k \operatorname{var}\left(z_k\right)/\tau_k^2 \tag{C5}$$

We finally obtain

$$D_i = \frac{1}{2}\sqrt{\frac{\pi}{\lambda}} \frac{T_G}{\Theta_G} \exp\left[2\lambda\Theta_G^2\right]\left\{\nu_i^2 \exp\left[2E(U_i)\right]\right\} \tag{C6}$$

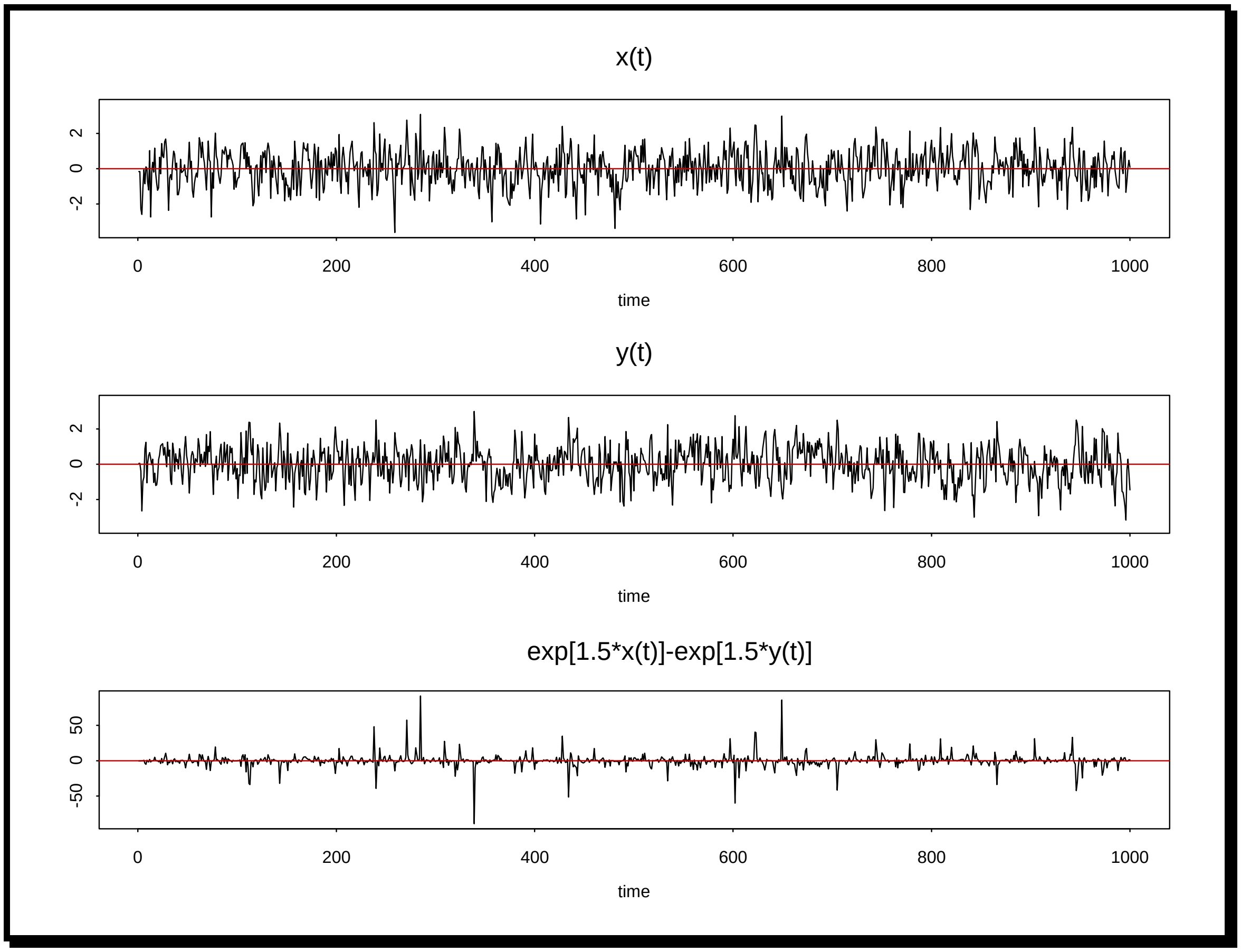

**Figure 1. Illustration of the notion of burstiness.**

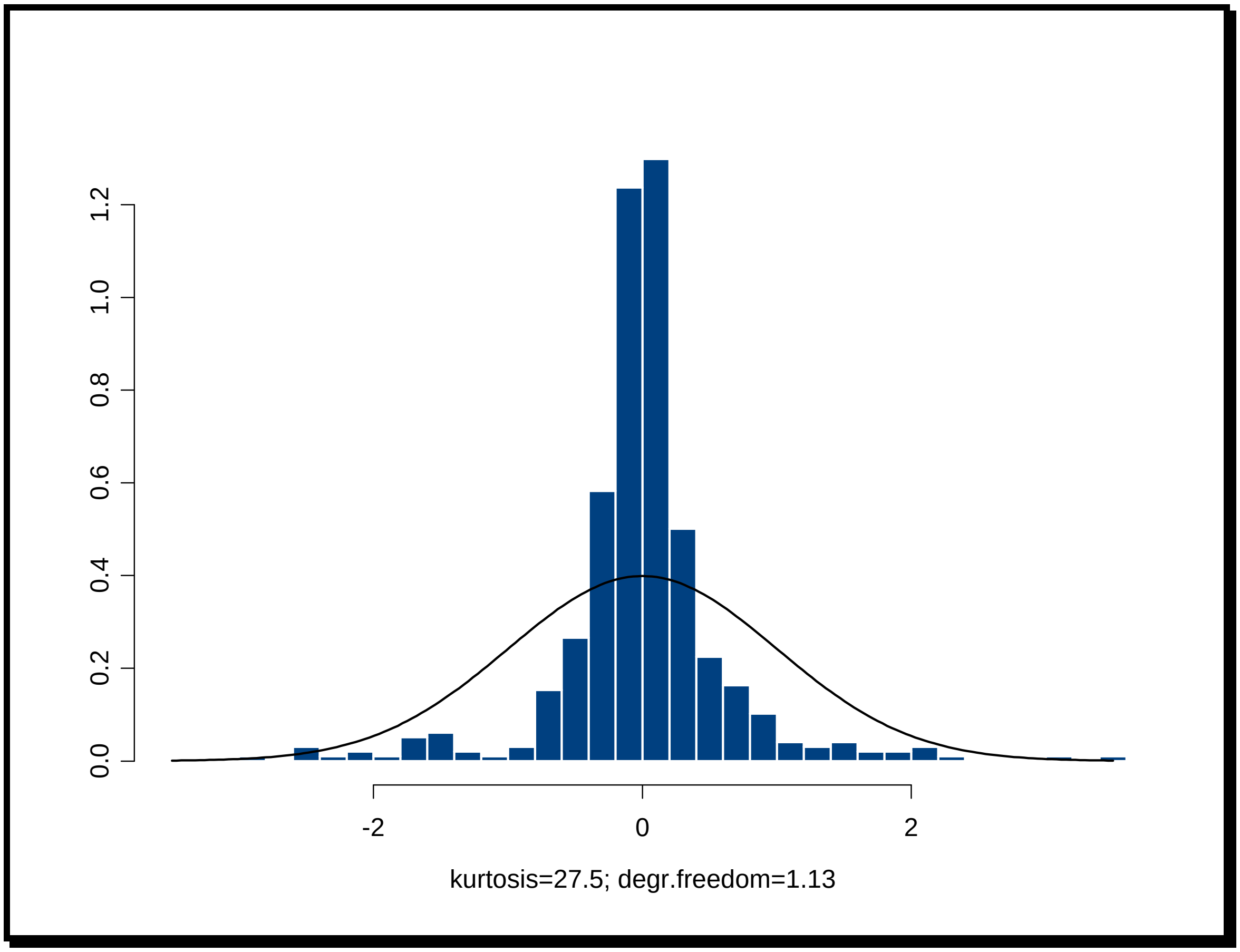

**Figure 2. Histogram of the process depicted in Figure 1. The distribution is close to the Students t with number of degrees of freedom 1.13. This is an indicator of "heavy tails". Solid line belongs to the standard normal distribution $N(0,1)$.**

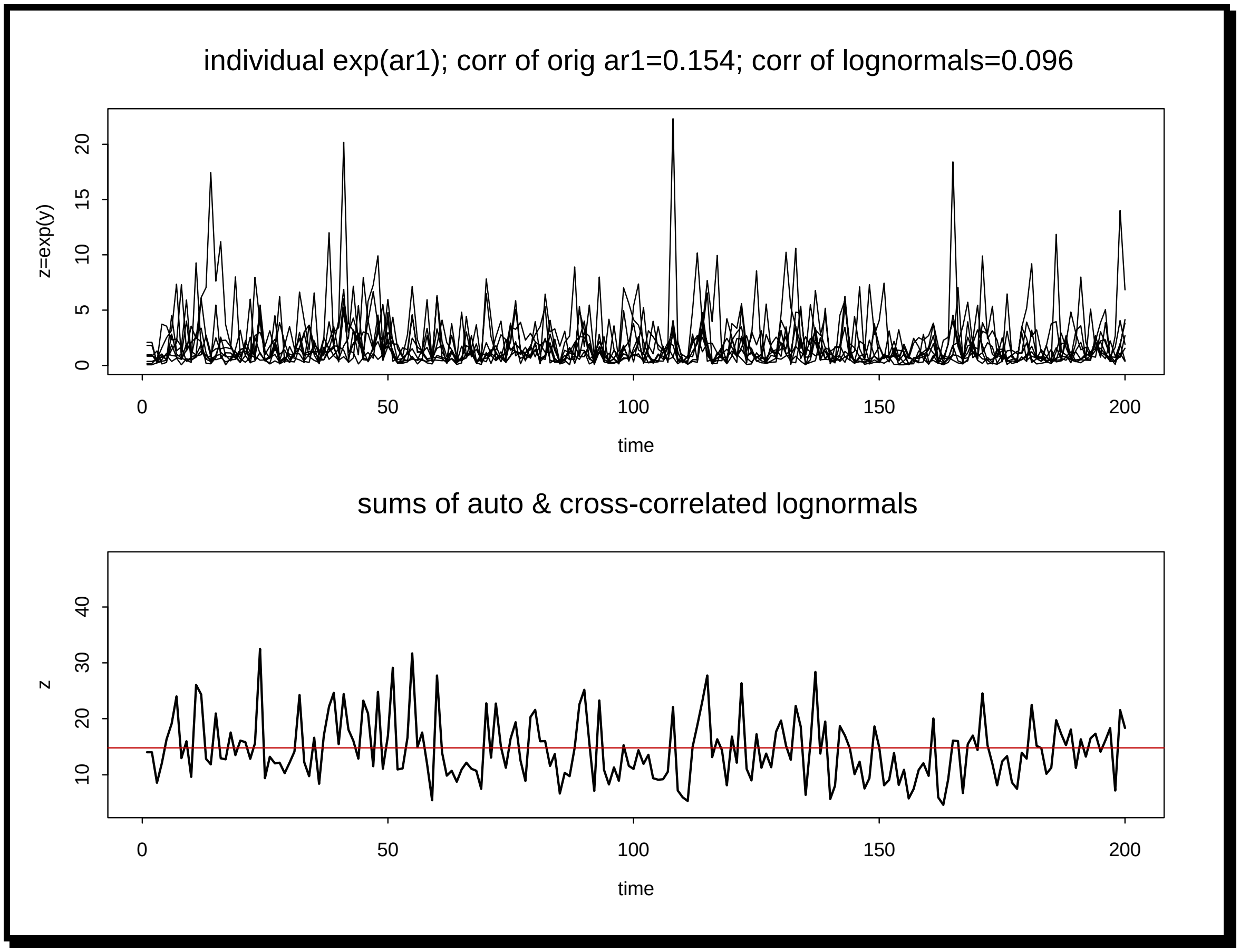

**Figure 3. Convergence of the sums of lognormal processes (top panel) to approximate normality (bottom panel).**

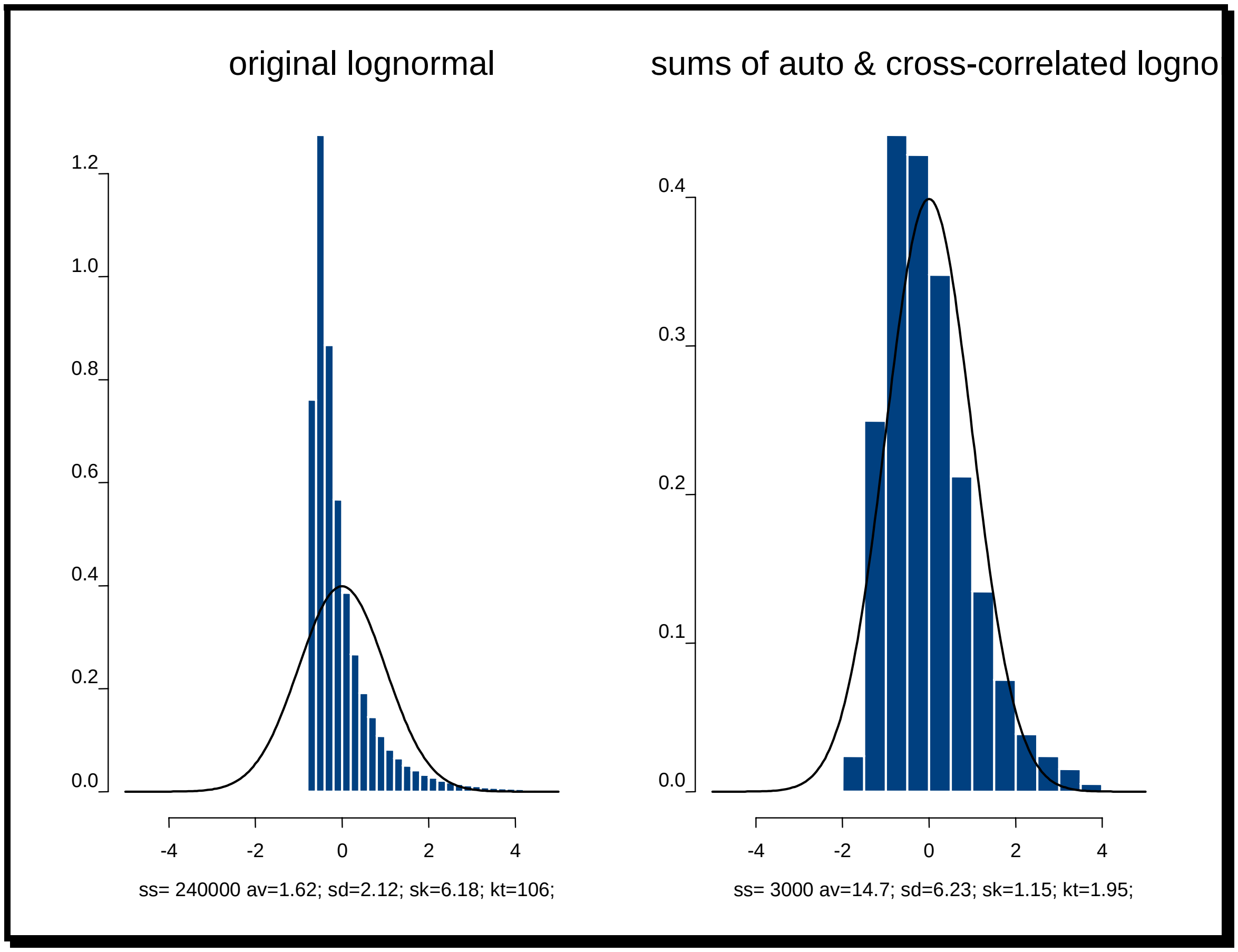

**Figure 4. Histograms of processes shown in Figure 3. Left panel: lognormal processes (skeweness 6.2, kurtosis 106). Right panel: distribution of sums of 80 lognormals (skeweness 1.2, kurtosis 2). In both cases, solid lines belong to standard normal.**

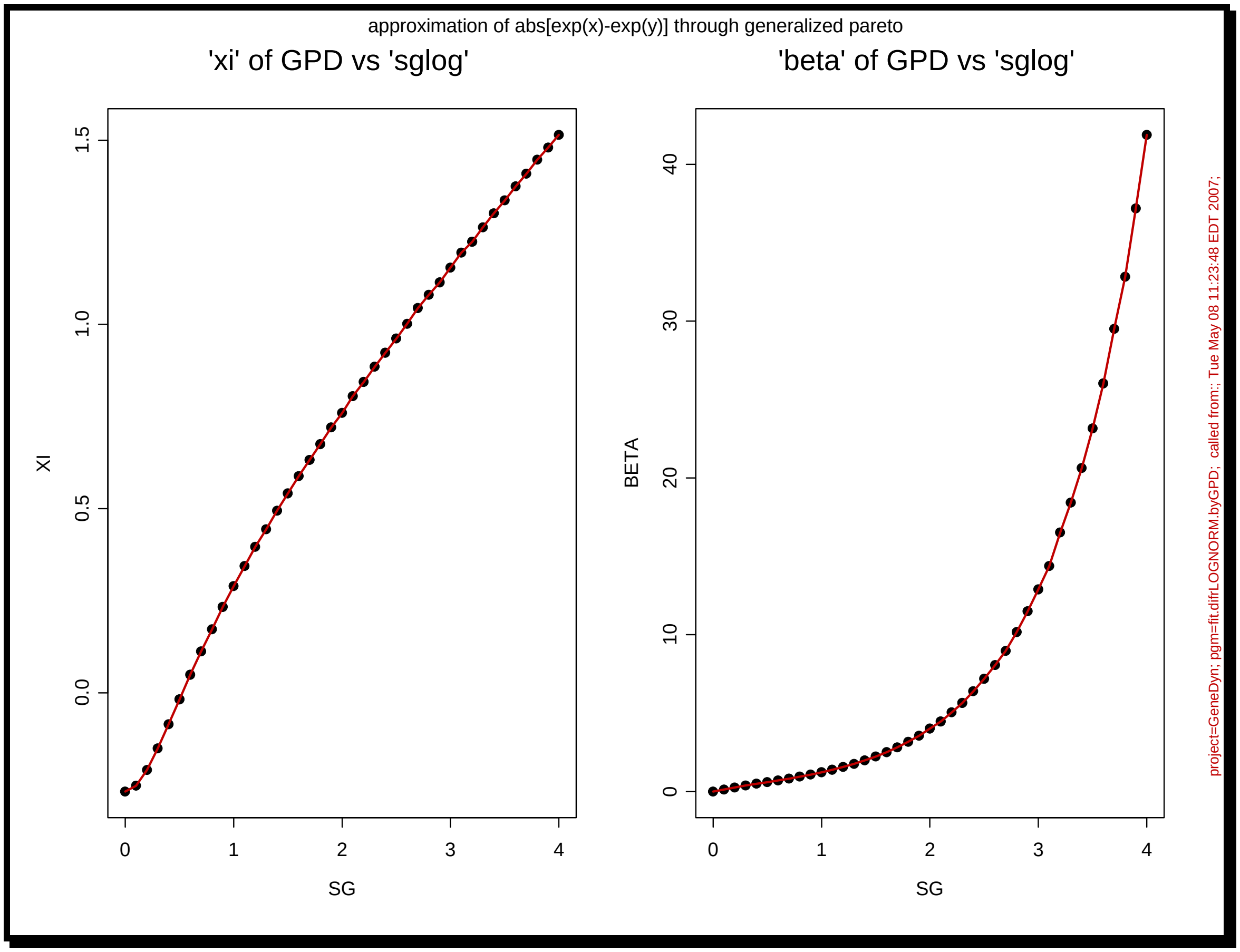

**Figure 5. Parameters of GPD expressed through the standard deviation, $\sigma$ . Dots are the parameters obtained by fitting the GPD to the simulated $h_\sigma = |\exp(\sigma x) - \exp(\sigma y)|$ ; solid lines are the parameters obtained through the analytical approximations (13-14).**

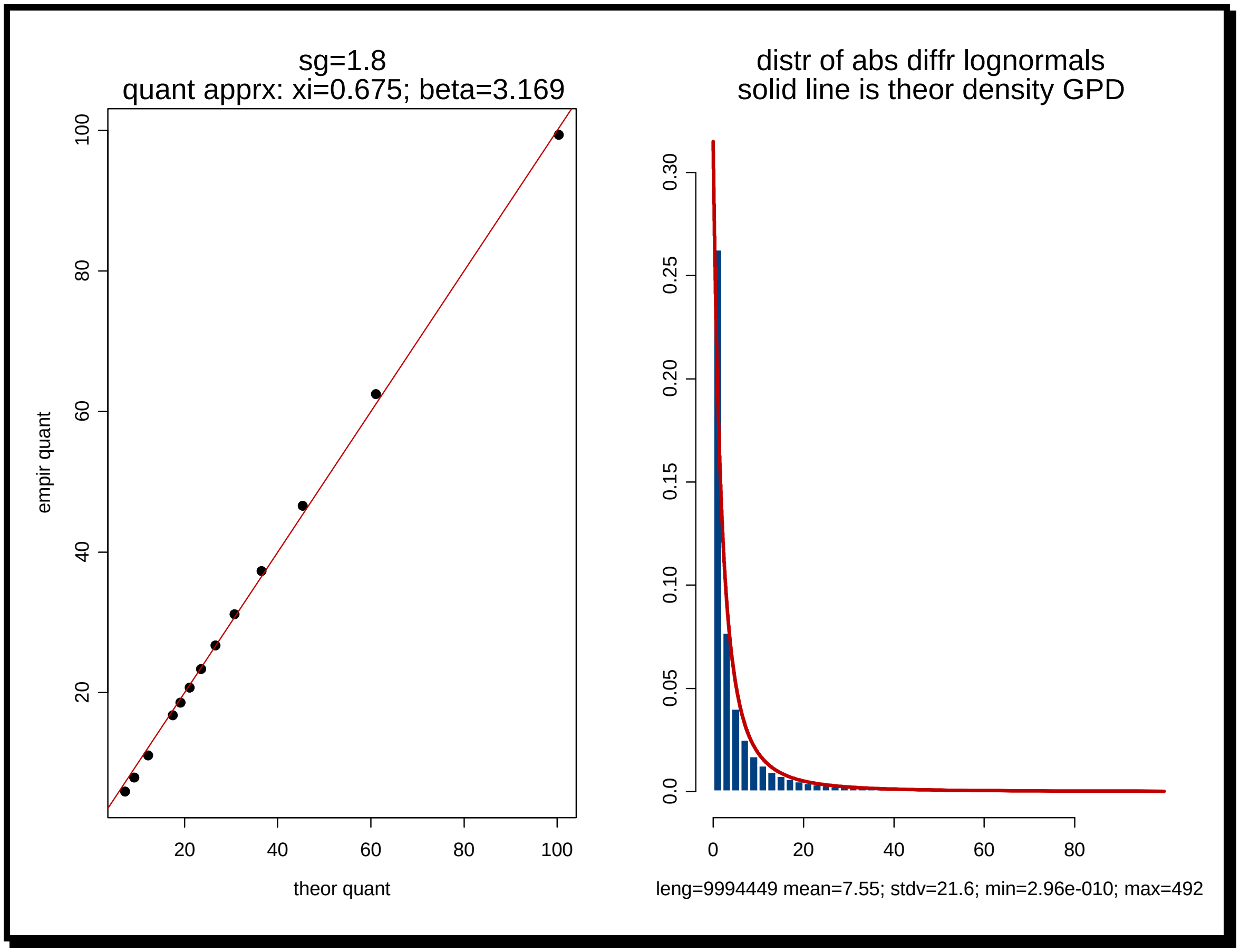

**Figure 6. Example of approximation of the difference of two lognormals by the GPD. Left panel: QQ-plot of theoretical GPD vs empirical $h_\sigma(t) = [\sigma x(t)] - \exp[\sigma y(t)]$; right panel: empirical histogram of $h_\sigma(t)$ vs. theoretical GPD density.**

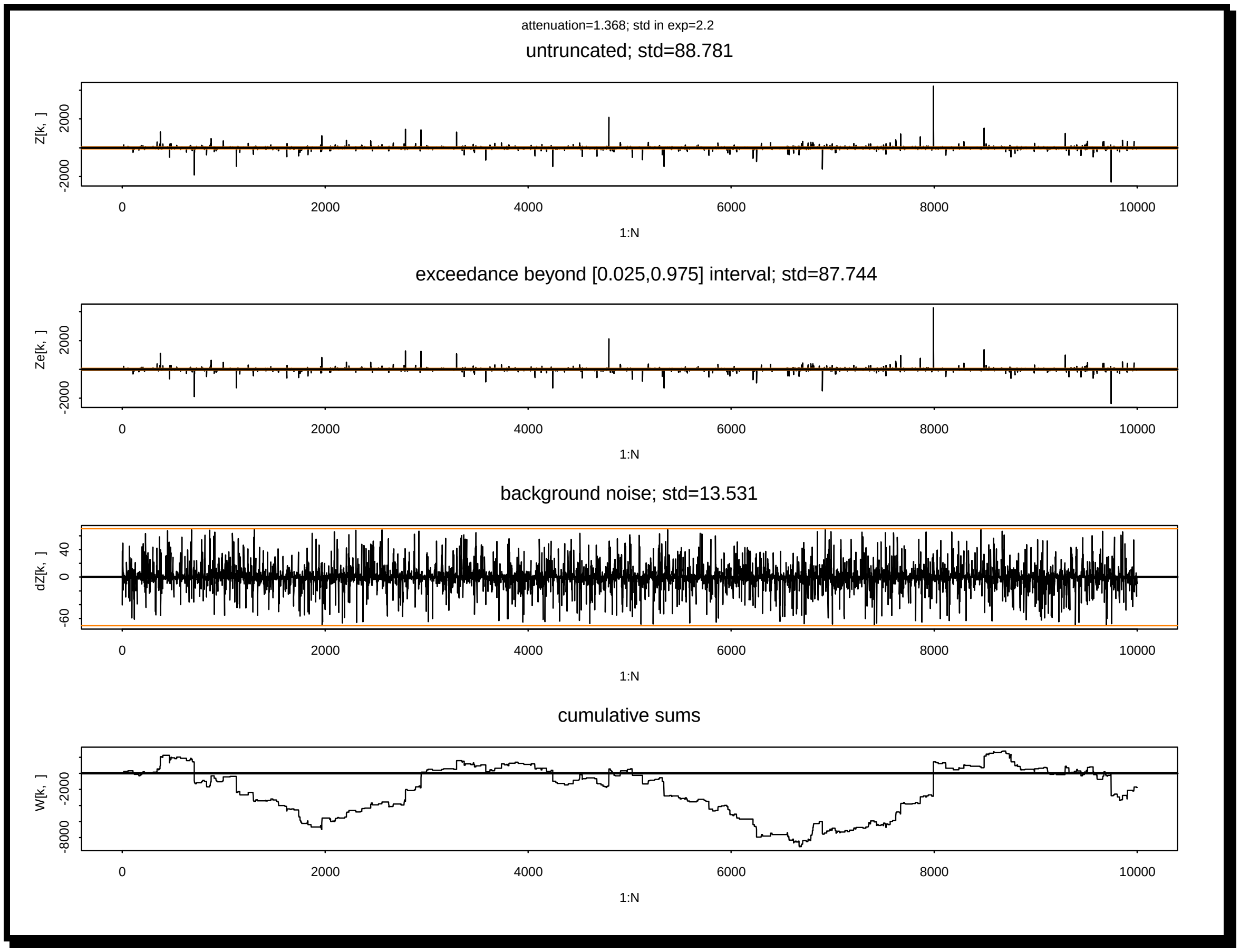

**Figure 7. Top panel: process $h_\sigma(t)$. Second panel: process of exceedances $\tilde{h}_\sigma(t)$. Third panel: residual noise, $d_\sigma(t) = h_\sigma(t) - \tilde{h}_\sigma(t)$. Bottom panel: trajectory of the random walk generated by $\tilde{h}_\sigma(t)$. Note that the variance of residual noise, $\mathrm{var}[d_\sigma(t)]$, is only 2.3% of total variance $\mathrm{var}[h_\sigma(t)]$, despite the fact that exceedances, $\tilde{h}_\sigma(t)$, occupy only 5% of the probability space.**

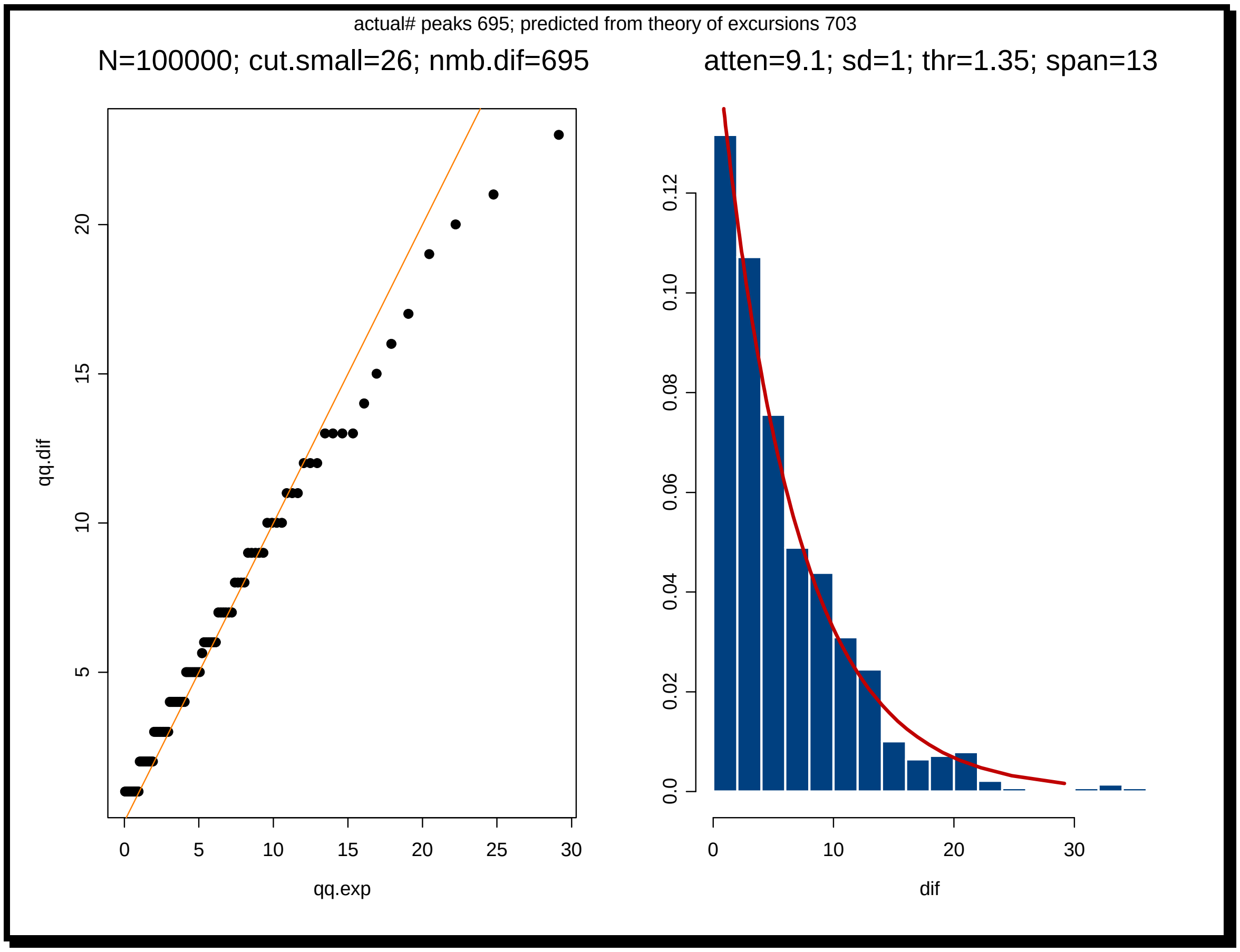

**Figure 8. Evidence that the exceedances form a Poisson process: waiting times are exponentially distributed.**

# Reference List

[1] T.Schlitt, A.Brazma, Modelling gene networks at different organisational levels, FEBS Lett. 579 (2005) 1859.

[2] P.Bork, et al., Protein interaction networks from yeast to human, Curr. Opin. Struct. Biol 14 (2004) 292.

[3] O.Fiehn, Metabolomics -- the link between genotypes and phenotypes, Plant Mol. Biol 48 (2002) 155.

[4] R.Raman, S.Raguram, G.Venkataraman, J.C.Paulson, R.Sasisekharan, Glycomics: an integrated systems approach to structure-function relationships of glycans, Nat. Methods 2 (2005) 817.

[5] G.Krauss, Biochemistry of Signal Transduction and Regulation, Wiley-VCH, 1999.

[6] I.Golding, E.C.Cox, RNA dynamics in live Escherichia coli cells, Proc. Natl. Acad. Sci. U. S. A 101 (2004) 11310.

[7] I.Golding, J.Paulsson, S.M.Zawilski, E.C.Cox, Real-time kinetics of gene activity in individual bacteria, Cell 123 (2005) 1025.

[8] M.Kaern, M.Menzinger, A.Hunding, A chemical flow system mimics waves of gene expression during segmentation, Biophys. Chem. 87 (2000) 121.

[9] J.Yu, J.Xiao, X.Ren, K.Lao, X.S.Xie, Probing gene expression in live cells, one protein molecule at a time, Science 311 (2006) 1600.

[10] A.Goldbeter, Computational approaches to cellular rhythms, Nature 420 (2002) 238.

[11] J.Paulsson, Summing up the noise in gene networks, Nature 427 (2004) 415.

[12] J.Paulsson, Prime movers of noisy gene expression, Nat. Genet. 37 (2005) 925.

[13] M.Scott, B.Ingalls, M.Kaern, Estimations of intrinsic and extrinsic noise in models of nonlinear genetic networks, Chaos. 16 (2006) 026107.

[14] M.A.Savageau, Biochemical systems analysis. I. Some mathematical properties of the rate law for the component enzymatic reactions, J. Theor. Biol 25 (1969) 365.

[15] M.A.Savageau, E.O.Voit, Recasting nonlinear differential equations as S-Systems, Math. Biosci. 87 (2006) 83.

[16] E.O.Voit, Canonical Nonlinear Modeling. S-System Approach to Understanding Complexity, Van Norstand Reinhold, NY, 1991.

[17] E.O.Voit, Computational Analysis of Biochemical Systems: A Practical Guide for Biochemists and Molecular Biologists, Cambridge University Press, Cambridge, UK, 2000.

[18] E.O.Voit, T.Radivoyevitch, Biochemical systems analysis of genome-wide expression data, Bioinformatics. 16 (2000) 1023.

[19] L.Tournier, Approximation of dynamical systems using S-Systems theory: Application to biological systems., International Symposium on Symbolic and Algebraic Computations, 2005, pp. 317-324.

[20] J.T.Kadonaga, Regulation of RNA polymerase II transcription by sequence-specific DNA binding factors (2004) .

[21] B.Lemon, R.Tjian, Orchestrated response: a symphony of transcription factors for gene control, Genes Dev. 14 (2000) 2551.

[22] S.Zumdahl, Chemical Principles, Houghton Mifflin, New York, 2005.

[23] G.Nicolis, I.Prigogine, Self-Organization in Nonequilibrium Systems: From Dissipative Structures to Order through Fluctuations, John Wiley & Sons, 1977.

[24] I.Prigogine, Time, Structure and Fluctuations. Nobel Lecture, 1977.

[25] F.R.Gantmacher, Applications of the Theory of Matrices, Interscience, NY, 1959.

[26] S.Rosenfeld, Stochastic Oscillations in Genetic Regulatory Networks, EURASIP Journal of Bioinformatics and Systems Biology (2006) 1.

[27] P.Nikolaev, D.Sokolov, Selection of an optimal biochemical reactor for microbiological synthesis, Chemical and Petroleum Engineering 16 (1980) 707.

[28] S.P.Bell, R.M.Learned, H.M.Jantzen, R.Tjian, Functional cooperativity between transcription factors UBF1 and SL1 mediates human ribosomal RNA synthesis, Science 241 (1988) 1192.

[29] M.Ptashne, Regulated recruitment and cooperativity in the design of biological regulatory systems, Philos. Transact. A Math. Phys. Eng Sci. 361 (2003) 1223.

[30] R.Bradley, Basic Properties of Strong Mixing Conditions. A Survey and Some Open Questions, Probability Surveys 2 (2005) 107.

[31] H.Cramer, R.Leadbetter, Stationary and Related Stochastic Processes, Wiley, NY, 1967.

[32] A.McNeil, R.Frey, P.Embrechts, Quantitative Risk Management, Princeton University Press, Princeton and Oxford, 2005.

[33] M.Loeve, Probability theory (The University series in higher mathematics), Van Nostrand, Holland, 1963.

[34] C.W.Gardiner, Handbook of Stochastic Methods: For Physics, Chemistry, and the Natural Sciences, Springer-Verlag, 1983.

[35] M.Kaern, M.Menzinger, A.Hunding, A chemical flow system mimics waves of gene expression during segmentation, Biophys. Chem. 87 (2000) 121.

[36] A.Goldbeter, Complex oscillatory phenomena, including multiple oscillations, in regulated biochemical systems, Biomed. Biochim. Acta 44 (1985) 881.

[37] J.Wu, N.Mehta, J.Zhang, A Flexible Lognormal Sum Approximation Method, Mitsubishi Electric Research Laboratories, 2005.

[38] J.Goutsias, N.H.Lee, Computational and experimental approaches for modeling gene regulatory networks, Curr. Pharm. Des 13 (2007) 1415.

[39] J.Lynch, On the threshold of chaos in random Boolean cellular automata, Random Structures and Algorithms 6 (1995) 239.

[40] T.Reil, Dynamics of Gene Expression in an Artificial Genome - Implications for Biological and Artificial Ontogeny, Advances in Artificial Life, 5th European Conference, ECAL'99, 1999, pp. 457-466.

[41] deJong H., Modeling and simulation of genetic regulatory systems: a literature review, J. Comput. Biol 9 (2002) 67.

[42] D.T.Gillespie, Stochastic Simulation of Chemical Kinetics, Annu. Rev. Phys. Chem. (2006) .

[43] L.Perko, Differential Equations and Dynamical Systems, Springer-Verlag, 2001.